\documentclass{nature}

\usepackage{graphicx}
\usepackage{amsmath}
\usepackage{amssymb}
\usepackage[colorlinks, citecolor=purple]{hyperref}
\usepackage[dvipsnames]{xcolor}
\usepackage{bm}
\usepackage{bbm}
\usepackage{caption}

\makeatletter
\let\saved@includegraphics\includegraphics
\AtBeginDocument{\let\includegraphics\saved@includegraphics}
\renewenvironment{figure}{\@float{figure}}{\end@float}
\makeatother

\usepackage{lineno}

\begin{document}
\title{Correlation measurement of propagating microwave photons at millikelvin}
\author{Aarne Ker{\"a}nen$^{1,\dag}$, Qi-Ming Chen$^{1,\dag,\ast}$, Andr{\'a}s Gunyh{\'o}$^{1}$, Priyank Singh$^{1}$, Jian Ma$^{1}$, Visa Vesterinen$^{2}$, Joonas Govenius$^{2}$, Mikko M{\"o}tt{\"o}nen$^{1,2,\ast}$}

\maketitle

\begin{affiliations}
 \item QCD Labs, QTF Centre of Excellence, Department of Applied Physics, Aalto University, FI-00076 Aalto, Finland.
 \item VTT Technical Research Centre of Finland Ltd. $\&$ QTF Centre of Excellence, P.O. Box 1000, 02044 VTT, Finland.\\
 $^{\dag}$These authors contributed equally to this work.\\
 $^{\ast}$Corresponding authors. E-mails: qiming.chen@aalto.fi (Q.C.); mikko.mottonen@aalto.fi (M.M.)\\
\end{affiliations}

\begin{abstract}
Microwave photons are important carriers of quantum information in many promising platforms for quantum computing.
They can be routinely generated, controlled, and teleported in experiments, indicating a variety of applications in quantum technology. However, observation of quantum statistical properties of microwave photons remains demanding:
The energy of several microwave photons is considerably smaller than the thermal fluctuation of any room-temperature detector, while amplification necessarily induces noise.
Here, we present a measurement technique with a nanobolometer that directly measures the photon statistics at the millikelvin temperature and overcomes this trade-off.
We apply our method to thermal states generated by a blackbody radiator operating in the regime of circuit quantum electrodynamics.
We demonstrate the photon number resolvedness of the nanobolometer, and reveal the $n(n+1)$-scaling law of the photon number variance as indicated by the Bose--Einstein distribution.
By engineering the coherent and incoherent proportions of the input field, we observe the transition between super-Poissonian and Poissonian statistics of the microwave photons from the bolometric second-order correlation measurement.
This technique is poised to serve in fundamental tests of quantum mechanics with microwave photons and function as a scalable readout solution for a quantum information processor.
\end{abstract}

Microwaves with the wavelength ranging from several millimeters to several meters are extensively used for communication, sensing, and power applications\,\cite{Pozar2011}. In the realm of quantum physics, many competitive platforms for quantum computation, such as superconducting quantum circuits\,\cite{Blais2021}, quantum dots\,\cite{Hanson2007}, and circular Rydberg atoms\,\cite{Raimond2001}, are also operated in this regime. These devices have to be cooled down to a cryogenic temperature where the quantum statistical properties of the microwave photons become prominent over the thermal noise. Weak microwave signals are subsequently amplified from the cryogenic to the room temperature stages for detection, which normally adds more than $10$ noise photons as referred to the input of a cryogenic high-electron-mobility transistor (HEMT) amplifier\,\cite{Menzel2010}. For many years, attempts have been made to circumvent the amplification noise by a variety of parametric processes\,\cite{Yurke1989, Yamamoto2008, CastellanosBeltran2008, Bergeal2010, Macklin2015, Bochmann2013, Bagci2014, Andrews2014, Forsch2019, Delaney2022}. An ideal degenerate parametric amplifier may be noiseless at a price of squeezing the quantum states in one field quadrature. Breaking the degeneracy solves the phase-nonpreserving problem, but it must add at least half of a quantum of noise to each quadrature of the input signal as required by quantum mechanics\,\cite{Clerk2010}. The quantum statistical properties of microwave photons can only be inferred from hours-long heterodyne detection of the noise itself and the amplified noisy signal, and through a complicating de-convolution procedure\,\cite{Eichler2012}. 

An alternative but yet to-be-demonstrated approach searches for a cryogenic detector that carries out the projective quantum measurements directly at the millikelvin temperature without amplification noise. In addition to the early attempts of bistable detectors\,\cite{Siddiqi2004}, to date the most successful protocol uses the two states of a qubit as the pointer\,\cite{Opremcak2018, Lescanne2020}. With a certain repetition rate, the qubit can be operated as a ``click" detector that gets excited when absorbing an incoming photon within a narrow frequency range. Recently, a superconductor-normal metal-superconductor (SNS) junction-based nanobolometer has been invented as a broad-band cryogenic microwave power detector with record-breaking speed and sensitivity\,\cite{Govenius2014, Govenius2016, Kokkoniemi2019, Kokkoniemi2020}. This nanobolometer requires no pulse sequences for operation and has the potential to be used as a photon-number resolved microwave photon detector. Here, we report a measurement technique using the nanobolometer to measure the quantum statistical properties of thermal microwave photons at the cryogenic temperature. 

\begin{figure}
  \centering
  \includegraphics[width=18cm]{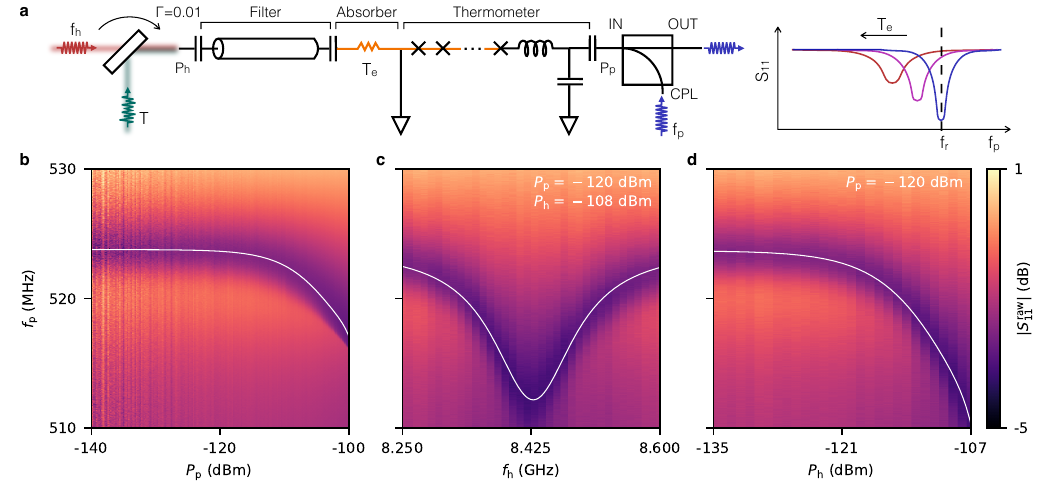}
  \linespread{1.2}
  \caption{{\bf Measurement scheme and characterization of the nanobolometer with coherent input.}
{\bf a} Schematic diagram of the experimental setup. A beam splitter with a transmission rate $\Gamma=0.01$ combines a coherent microwave field at frequency $f_{\rm h}$ with a thermal field characterized by the radiation temperature $T$. The transmitted field is filtered and subsequently absorbed by a resistive absorber, causing a change of the electron temperature, $T_{\rm e}$, in the normal-metal nanowire. It leads to a decrease of the critical current of the SNS junctions (crosses) and hence a down-shift in the resonance frequency of the thermometer. This shift is measured by a weak probe field at frequency $f_{\rm p}$, which is coupled to the thermometer via a directional coupler (see Extend Data Fig.\,1).
{\bf b} Reflection magnitude of the thermometer as a function of the probe power at the sample, $P_{\rm p}$, and the probe frequency, $f_{\rm p}$. The solid curve represents a cubic fit of the resonance frequency. 
{\bf c} Reflection magnitude of the thermometer as a function of the input frequency, $f_{\rm h}$, and the probe frequency. The probe and input powers are fixed at $P_{\rm p} = -120\,{\rm dBm}$ and $P_{\rm h} = -108\,{\rm dBm}$, respectively. The solid curve is a Lorentzian fit of the resonance frequency, which indicates a passband of ${\rm FWHM} = 133\,{\rm MHz}$ around the central frequency $f_{0}=8.428\,{\rm GHz}$.
{\bf d} Reflection magnitude of the thermometer as a function of the input power at the sample, $P_{\rm h}$, and the probe frequency, $f_{\rm p}$. Here, the probe power is fixed at $P_{\rm p} = -120\,{\rm dBm}$ and $f_{\rm h}$ is set to the center frequency of the filter. The solid curve is a cubic fit of the resonance frequency. The data is normalized to $0\,{\rm dB}$ at $530\,{\rm MHz}$ probe frequency in (b)--(d).
}
\end{figure}

The nanobolometer, as shown in Fig.\,1a, has a transmission-line filter that selects the bandwidth of interest. The filter is terminated by a normal-metal nanowire with an approximately $50\,{\Omega}$ resistance, i.e., the absorber. The absorber transforms the electromagnetic energy of the incident photons to a significant increase of the electron temperature in the nanowire. This nanowire extends through several superconducting islands and forms an array of short SNS junctions that are embedded in a sub-gigahertz probe resonator, i.e., the thermometer\,\cite{Chen2023b}. The hot electrons weaken the coherent Andreev reflections at the SN and NS interfaces of each junction, and cause a down-shift of the mean resonance frequency of the thermometer. The amount of the shift is determined by the mean absorbed power, which for a given frequency and bandwidth can be characterized by the mean photon number of the input field, $\langle \hat{n}\rangle$, considered in the unit of per second per hertz as it is the conventional unit for describing a stationary beam of electromagnetic field (see Supplementary Note 1). The broadening of the line-shape is related to the photon number variance, $(\Delta n)^2 = \langle \hat{n}^2 \rangle-\langle \hat{n} \rangle^2$, similar to the Doppler broadening in laser spectroscopy\,\cite{Demtroeder1996} (see Methods). The nanobolometer therefore allows direct readout of the first two orders of photon number moments, $\langle \hat{n}\rangle$ and $\langle \hat{n}^2\rangle$, from the long-time-averaged scattering response of the thermometer without any amplification of the input field. Note that the incident and probe fields are not directly electrically coupled to each other, owing to the ground port between the absorber and the thermometer, and have several gigahertz difference in frequency.

Here, we use a cryogenic blackbody radiator to generate thermal microwave photons for bolometric measurement (see Methods). By controlling the local temperature of the radiator, $T$, we achieve an in-situ control of the mean radiation photon number according to the Planck's law, $\langle \hat{n}\rangle=1/\left\{\exp\left[hf_{\rm h}/(k_{\rm B}T)\right]-1\right\}$, where $f_{\rm h}$ is the input photon frequency, and $h$ and $k_{\rm B}$ are the Planck and Boltzmann constants, respectively. With the decrease of $\langle \hat{n}\rangle$, the photon number variance converges from a quadratic function, $(\Delta n)^2\approx\langle \hat{n}\rangle^2$ at $\langle \hat{n}\rangle \gg 1$, to a linear relation, $(\Delta n)^2\approx\langle \hat{n}\rangle$ at $\langle \hat{n}\rangle \ll 1$\,\cite{Goetz2017}. Correspondingly, the photon statistics crosses over the boundary between the classical and the quantum regimes of the radiation field and converges to the shot-noise limit, indicating the discreteness of photons. This process is precisely described by the Bose-Einstein distribution of indistinguishable bosons, where $(\Delta n)^2=\langle \hat{n}\rangle \left(\langle \hat{n}\rangle+1 \right)$. 

We also use a weak coherent input field as the reference of the bolometric measurement, where $(\Delta n)^2=\langle \hat{n}\rangle$ by definition. Despite the different scaling laws of $(\Delta n)^2$, a stark contrast between the thermal and coherent photons lies in the second-order correlation function, $g^{(2)}(\tau)$, with $\tau$ being the time delay. It characterizes the temporal separation between two successive photons in a propagating field. At zero delay, we have $g^{(2)}(0) =1 + \left[(\Delta n)^2 -\langle \hat{n} \rangle \right]/\langle \hat{n} \rangle^2$. A perfect coherent field obeys Poissonian statistics with $g^{(2)}(0)= 1$, while thermal photons are super-Poissonian with $g^{(2)}(0)= 2$. Considering that $g^{(2)}(\tau) = 1$ for $\tau \rightarrow \infty$, an observation of $g^{(2)}(0)> 1$ indicates the photon bunching effect where the photons tend to propagate in bundles. 

\section*{Results}
\subsection{Characterization of the nanobolometer.}
In our experiment, we first sweep the probe frequency from $f_{\rm p}=510$ to $530\,{\rm MHz}$ to identify the resonance frequency of the thermometer (Fig.\,1b). At the minimum probe power, $P_{\rm p}=-140\,{\rm dBm}$ defined at the thermometer input by assuming an $80\,{\rm dB}$ attenuation from the source, we extract the resonance frequency as $f_{\rm r} = 524\,{\rm MHz}$, as well as the external and the total energy decay rates $\gamma_{\rm c} = 4.8\,{\rm \mu s^{-1}}$ and $\gamma = 18.7\,{\rm \mu s^{-1}}$, respectively, using the circle-fit method\cite{Chen2022}. We observe that $f_{\rm r}$ remains almost invariant at $P_{\rm p}<-110\,{\rm dBm}$, but it starts to bend downwards at higher probe power together with the line-shape changing from a symmetric into an asymmetric form. Possible sources of this non-linearity are the coherent and incoherent Andreev reflections in the short SNS junctions, leading to the Josephson effect\,\cite{Chen2023} and electrothermal feedback\,\cite{Visser2010, Govenius2016}, respectively. 

We then set the probe power at $P_{\rm p}=-120\,{\rm dBm}$, and characterize the filter passband with a coherent input signal. Here, we sweep the input frequency from $f_{\rm h}=8.25$ to $8.60\,{\rm GHz}$ with a fixed power $P_{\rm h}=-108\,{\rm dBm}$ defined at the filter input by fitting an $88\,{\rm dB}$ attenuation from the source (Fig.\,1c). Depending on the frequency of the input signal as compared to the passband, the resonance frequency of the thermometer, $f_{\rm r}$, shifts as a Lorentzian function of $f_{\rm h}$. The center frequency and bandwidth of the filter are determined to be $f_{0}=8.428\,{\rm GHz}$ and ${\rm FWHM}=133\,{\rm MHz}$, respectively.

Subsequently, we fix the input frequency at the center of the passband, $f_{\rm h} = f_{0}$, and vary its power from $P_{\rm h}=-135$ to $-107\,{\rm dBm}$ (Fig.\,1d). The other parameters are kept identical to those in Fig.\,1c. We observe that $f_{\rm r}$ decreases with increasing $P_{\rm h}$, demonstrating the functionality of the nanobolometer as a power detector. In contrast to Fig.\,1b, the line-shape remains symmetric owing to a different mechanism than the nonlinearity. Here, the shift is caused by the increase of the electron temperature, $T_{\rm e}$, in the short SNS junctions, leading to an increase of the inductance in the resonant circuit. The relative shift, $\Delta f_{\rm r}$, compared with the highest value of $f_{\rm r}$ can be well fitted by a cubic function of $P_{\rm h}$. Here, the input photon number, $\langle \hat{n} \rangle = P_{\rm h}/({\rm FWHM}h f_{\rm 0})$, is estimated to be $0$--$27\,{\rm photons/(s \times Hz)}$ (see Extended Data Fig.\,2 for cryogenic setup). 

\subsection{Photon number resolution.}
\begin{figure}
  \centering
  \includegraphics[width=9cm]{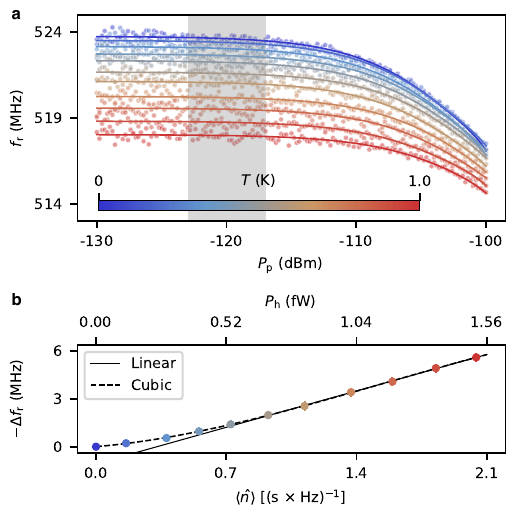}
  \linespread{1.2}
  \caption{{\bf Photon-number resolved frequency shift.} 
{\bf a} Resonance frequency of the thermometer, $f_{\rm r}$, as a function of the probe power, $P_{\rm p}$, at different indicated temperatures of the blackbody radiation, $T$.
The dots and solid curves are the corresponding experimental data and cubic fits, respectively.
{\bf b} Frequency shift, $\Delta f_{\rm r}$, in the region highlighted with grey in (a) as a function of the mean photon number of the thermal radiation field, $\langle \hat{n} \rangle$. The dots and error bars represent the mean values and standard deviations of the raw data in the selected range. The dashed black line shows a cubic fit to the full data and the solid black line represents a linear fit to the data where $\langle \hat{n} \rangle \geq 1$. The top horizontal axis indicates the radiation power in the entire $133\,{\rm MHz}$ bandwidth of the input filter.
}
\end{figure}

For a more precise benchmark of the frequency shift with different $\langle \hat{n} \rangle$, we use a blackbody radiator as the nanobolometer input. We sweep the radiation temperature, $T$, from the base temperature to approximately $1\,{\rm K}$ while keeping the mixing chamber (MXC) temperature below $60\,{\rm mK}$. The corresponding input photon number, $\langle \hat{n} \rangle$, can be precisely determined by the Planck's law, which varies from $0$ to $2\,{\rm photons/(s \times Hz)}$. At each radiation temperature, we sweep both $P_{\rm p}$ and $f_{\rm p}$ and identify the resonance frequency, $f_{\rm r}$, at the dip of the averaged reflection magnitude. Here, the intermediate-frequency (IF) bandwidth is set to $200\,{\rm Hz}$ with $10$ averages. Figure\,2a summarizes the measurement results. A clear decrease of $f_{\rm r}$ with increasing $T$ is observed, which is consistent with the observation with a coherent input field as shown in Fig.\,1d. 

Averaging the relative frequency shift within a $\pm 3\,{\rm dB}$ range around $P_{\rm p}=-120\,{\rm dBm}$, we observe a cubic relation between $\Delta f_{\rm r}$ and $\langle \hat{n} \rangle$ (Fig.\,2b). At the minimum value of $\langle \hat{n} \rangle=0.16\,{\rm photons/(s\times Hz)}$, we obtain the mean and standard deviation of $-\Delta f_{\rm r}$ as $0.22\,{\rm MHz}$ and $0.19\,{\rm MHz}$, respectively, corresponding to a coefficient of variation (CV) less than $1$. We therefore attribute $0.16\,{\rm photons/(s\times Hz)}$ as an upper bound of the photon number resolvedness of the nanobolometer. The value of CV decreases monotonically with $\langle \hat{n} \rangle$ and approaches $0.05$ at the highest photon number. In fact, this resolution may be pushed to an even smaller value by decreasing the IF bandwidth or increasing the number of averages. 

We note that the relatively small photon number resolution is benefited from the broad bandwidth of the blackbody radiation that covers the entire $133\,{\rm MHz}$ passband of the input filter. The corresponding input-field-independent power resolution of the nanobolometer is $119\,{\rm aW}$, corresponding to a considerable amount of individual photon wavepackets per second, $2.1\times 10^{7}\,{\rm photons/s}$. In this regard, the nanobolometer reported here is still far from resolving each individual photon emission event in a microwave field, which has been demonstrated in the optical regime \,\cite{Wiersig2009}. Fortunately, the observed power resolution is adequate for resolving the quantum statistical properties of propagating microwave photons that are conventionally defined by the arriving photons per second per hertz\,\cite{Goetz2017} (see Supplementary Note 1).

\subsection{Photon number variance.}
\begin{figure}
  \centering
  \includegraphics[width=9cm]{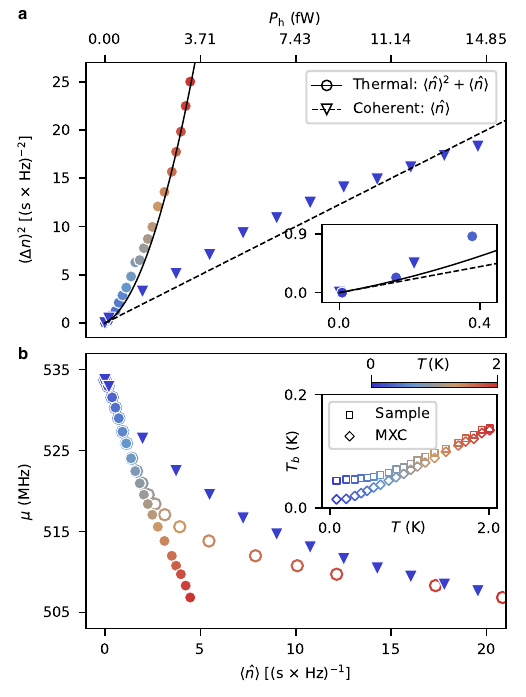}
  \linespread{1.2}
  \caption{{\bf Photon number variance extracted from the reflection spectrum.} 
  {\bf a} Photon number variance, $(\Delta n)^2$, as a function of the mean input photon number, $\langle \hat{n} \rangle$, for the thermal (circles) and coherent (triangles) input fields. The solid and dashed lines represent the theoretical expectation for thermal and coherent states, respectively. The thermal photon number is obtained by Planck's law, and the coherent photon number is obtained by assuming an $88\,{\rm dB}$ attenuation in the input line as a scaling parameter. The detection efficiency of the nanobolometer is assumed to be unity to minimize the number of fitting parameters. The inset shows the region near the origin. 
  {\bf b} Extracted resonance frequency, $\mu$, as a function of $\langle \hat{n} \rangle$ for thermal (circles) and coherent (triangles) input fields. The open circles are corrected results by considering the phonon temperature change, $T_{\rm b}$, during the temperature sweep. The inset shows the records of two different sensors that are mounted at the sample holder and the MXC, respectively. The former is lower-bounded by approximately $45\,{\rm mK}$ due to self-heating. The top horizontal axis indicates the radiation power in the entire $133\,{\rm MHz}$ bandwidth of the input filter.
}
\end{figure}

The quantum statistical properties of thermal states are fully determined by the mean photon number, $\langle \hat{n} \rangle$, and its variance, $(\Delta n)^2$. With a weak radio-frequency (RF) probe field applied to the thermometer, these two values can be respectively obtained from the fitted resonance frequency, $\mu$, and the broadening of the spectrum linewidth, $\sigma^2$ (see Methods). Here, we sweep the radiation temperature, $T$, from the base temperature to approximately $2\,{\rm K}$ and measure the averaged reflection coefficient, $S_{11}^{\rm ave}$, at a fixed probe power, $P_{\rm p}=-120\,{\rm dBm}$. We first fit the theoretical curve to the data at the minimum $T$ as a reference curve for zero broadening ($\sigma^2 = 0$). We then allow $\sigma^2$ to be a fitting parameter and extract its value at different $T$ by comparing the measurement results with this reference (see Supplementary Note 4). The photon number variance is obtained by assuming a linear relation $\Delta n = \alpha \sigma$, where $\alpha=1.97\,{\rm photons/MHz}$ is a fitted scaling factor.

At a sufficiently small photon number, where $\langle \hat{n} \rangle$ is precisely determined by the Planck's law, the extracted photon number variance, $(\Delta n)^2$, is almost equal to $\langle \hat{n} \rangle$, as shown in the inset of Fig.\,3a. It reveals the particle nature of photons in the form of shot noise, as described by the Poisson distribution. On the other hand, $(\Delta n)^2$ can be well approximated by a quadratic function, $\langle \hat{n} \rangle^2$, at large $\langle \hat{n} \rangle$. These two limits are connected by the Bose-Einstein distribution, which predicts $(\Delta n)^2\approx\langle \hat{n}\rangle \left(\langle \hat{n}\rangle+1 \right)$ for indistinguishable bosons in thermal states. Our experimental results in Fig.\,3a show an excellent agreement with the theoretical expectation. It reveals the transition between the quantum and classical regimes of a microwave field with the increase of the photon number.

For comparison, we replace the thermal input field by a coherent signal at $f_{0}$. We sweep the input power from $P_{\rm h} = -128$ to $-108\,{\rm dBm}$, and extract $\langle \hat{n} \rangle$ and $(\Delta n)^2$ with almost identical fitting parameters as for the thermal fields (see Supplementary Note 4). We observe a qualitatively different statistics of photons between the thermal and coherent input fields, as shown in Fig.\,3a. Here, $(\Delta n)^2$ remains as a pure linear function of $\langle \hat{n} \rangle$ in the entire range of the power sweep, corresponding to a mean photon number from $0$ to $19$. This observation is in agreement with the Poisson statistics of the input photons in a coherent state. The excellent agreement between theory and experiments for both thermal and coherent states with almost the same fitting parameter demonstrate the correctness of our measurement result.

Figure\,3b shows the extracted value of $\mu$ for both thermal and coherent inputs. The data are in quantitative agreement with each other for $\langle \hat{n} \rangle < 2$, indicating an almost linear relation between the frequency shift of the thermometer and the input power as observed in Fig.\,2b. However, a noticeable discrepancy is observed at the higher input power. Since the temperature of the blackbody radiator, $T$, is significantly higher than the MXC temperature in this range, we attribute the observed discrepancy to the heating of the phonon bath temperature, $T_{\rm b}$. The latter is recorded in the inset of Fig.\,3b. We then consider a $6^{\rm th}$-order polynomial correction of the net heating power due to the phonon temperature change, i.e., $P=\beta T_{\rm b}^{6} + P_{\rm h}$, for ultra-thin films \cite{Wei2008}. A qualitative agreement of $\mu$ for the two types of photons is achieved with $\beta  = 1.84\,{\rm nW/K^6}$, as shown in Fig.\,3b. A quantitative agreement can be achieved by assuming an approximately $50\%$ absorption rate of the nanobolometer, which is not taken into account throughout this study.

\subsection{Photon correlation.}
\begin{figure}
  \centering
  \includegraphics[width=9cm]{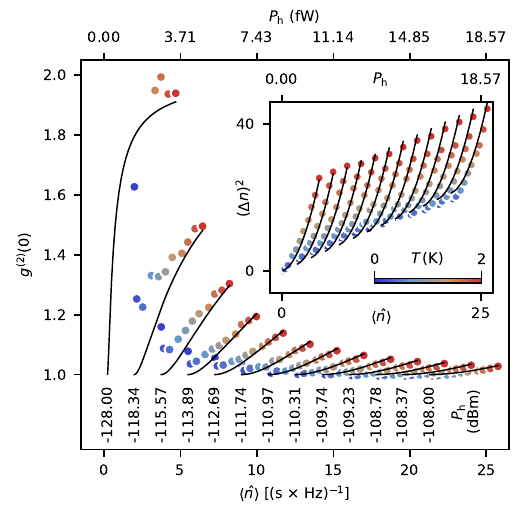}
  \linespread{1.2}
  \caption{{\bf Second-order correlation for different photon states.} 
The proportions of the coherent and thermal parts in the input field are controlled by varying the power of the coherent field, $P_{\rm h}$, and the radiation temperature, $T$. Here, the solid curves correspond to the theoretical expectations at fixed values of $P_{\rm h}$ but with a varying $T$. 
The inset shows the raw data of the photon number and variance that are used to calculate $g^{(2)}(0)$. The top horizontal axis indicates the radiation power in the entire $133\,{\rm MHz}$ bandwidth of the input filter.
}
\end{figure}

The shown ability of resolving the mean photon number, $\langle \hat{n} \rangle$, and its variance, $(\Delta n)^2$, indicates a convenient way of measuring the zero-delay second-order correlation functions of the incident microwave field, $g^{(2)}(0)$. In our experiment, we control the coherent and incoherent proportions of the input field by adjusting their powers at the beam splitter inputs (Fig.\,1a), and repeat the photon variance measurement as shown in Fig.\,3. 

Figure\,4 summarizes the correlation measurement result as well as the raw data of $\langle \hat{n} \rangle$ and $(\Delta n)^2$. At a fixed power of the coherent field, $P_{\rm h}$, the correlation function, $g^{(2)}(0)$, increases with the radiation temperature, $T$, from $1$ to a value slightly below $2$. It reveals the loss of coherence as the thermal proportion of the field increases. On the other hand, $g^{(2)}(0)$ converges to $1$ with increasing $P_{\rm h}$. This is consistent with our expectation since an ideal coherent field leads to $g^{(2)}(0)=1$. The measured value of $g^{(2)}(0)$ is larger than the theoretical expectation by a few percent where the coherent photon number is below $1\,{\rm photons/(s\times Hz)}$. We attribute this loss of coherence to the $88\,{\rm dB}$ attenuation in the input line, where the blackbody radiation at different thermalization stages necessarily impinges on the input signal. Overall, the observed $g^{(2)}(0)$ shows a quantitative agreement with the theoretical expectation. It is directly obtained from the resonance frequency and linewidth of the averaged scattering response without amplification and the corresponding noise. This is the key advantage of using cryogenic detectors for microwave photon correlation measurement compared with the combination of quantum limited amplifiers and room-temperature linear detectors\,\cite{Bozyigit2010, Lang2013}. 

\section*{Discussion}
The bolometric measurement of the second-order photon correlation function demonstrates the ability of using cryogenic detectors to characterize the quantum statistical properties of microwave radiation. This method implements projective quantum measurements directly at the millikelvin stage, and therefore only requires a data bus from the cryogenic to the room temperature stages to transfer the classical measurement outcome. Multiple nanobolometers can be multiplexed with the same bus \cite{Singh2024}. The combination of long and short SNS junctions allows the RF operation of the nanobolometer, which plays a key role in reading out the second-order photon number moments. This configuration also pushes the bandwidth of cryogenic detectors to the convenient sub-$10\,{\rm GHz}$ regime as compared to other designs such as transition edge sensor (TES)\,\cite{Cabrera1998}, kinetic inductance detector (KID)\,\cite{Day2003}, and superconducting nanowire single photon detector (SNSPD)\,\cite{Goltsman2001}. By combining this technique with correlation spectroscopy\,\cite{Demtroeder1996}, it is possible to measure even higher-order correlation functions of microwave photons with a single nanobolometer. An alternative upgrade for higher-order correlation measurements is to combine the nanobolometers with microwave beam splitters\,\cite{Mariantoni2010} in a similar way of the well established optical experiment setup. 

In the future, the capability of microwave correlation measurements may advance the current understandings of the cosmic microwave background (CMB) \cite{Bernardis2000} and resolve a possible distortion of the CMB frequency spectrum from a pure blackbody radiator. In addition, the small physical footprint and ultralow power consumption of the nanobolometer \cite{Kokkoniemi2019,Kokkoniemi2020} make it suitable for millikelvin integration. Integrating high-performance microwave photon generators \cite{Yan2021} and detectors \cite{Gunyho2024} at millikelvin enables an isolated cryogenic environment for quantum information processors. It opens up new possibilities of controlling and reading out the quantum information carried by microwave photons with less noise, latency, and power consumption.

\makeatletter
\setcounter{figure}{0}
\renewcommand{\figurename}{\textbf{Extended Data Fig.}}
\renewcommand{\thefigure}{\arabic{figure}}
\renewcommand{\fnum@figure}{\textbf{\figurename\,\thefigure}}
\makeatother

\section*{Methods}
\subsection{Sample preparation.}
The nanobolometer used in this study consists of an $7.5\,{\rm mm}$-long Nb coplanar waveguide filter, an $150\,{\rm nm}$-wide $1\,{\rm \mu m}$-long Au$_{3}$Pd absorber, and an $7$-island Al-Au$_{3}$Pd-Al junction thermometer (see Extended Data Fig.\,1)\,\cite{Govenius2016, Kokkoniemi2019}. The superconducting islands are $300\,{\rm nm}$ wide and $2.4\,{\rm \mu m}$ long, and they are evenly spaced by a $300\,{\rm nm}$ gap. The designed thicknesses of the Si substrate and the SiO${}_2$, Nb, Au$_{3}$Pd, and Al layers are $675\,{\rm \mu m}$, $300\,{\rm nm}$, $200\,{\rm nm}$, $30\,{\rm nm}$, and $100\,{\rm nm}$, respectively. 

The blackbody radiator uses an $20\,{\rm dB}$ attenuator as a microwave beam splitter with an $\Gamma=0.01$ transmission rate of the coherent signal and $1-\Gamma$ of the thermal radiation. It is tightly integrated with an $100\,{\Omega}$ resistor and a RuO$x$ sensor for local temperature control\,\cite{Goetz2017}. A weak thermal link between the radiator and the mixing chamber plate is constructed by a $2\,{\rm mm}$-wide and $300\,{\rm mm}$-long Cu braid. A dynamical balance between cooling and heating is achieved with a commercial PID temperature controller. We observe that the MXC temperature remains below $60\,{\rm mK}$ when the radiation temperature is below $1\,{\rm K}$. A sensor mounted directly at the sample holder indicates a slightly higher sample temperature, but it remains below $80\,{\rm mK}$ for $T < 1\,{\rm K}$ (see inset of Fig.\,3 for recorded temperatures of the two sensors).

\subsection{Experimental setup.}
The cryogenic and room-temperature setup for the experiment is shown in Extended Data Fig.\,2. We use a directional coupler at the thermometer input for reflection-type measurement, where the input signal is attenuated by approximately $20\,{\rm dB}$ but almost all the signal reflected from the sample is routed to the output port. The signal is subsequently amplified and down converted to an intermediate frequency of $62.5\,{\rm MHz}$ for measurement. Here, the filters before and after the mixers are used to remove the red sideband. The sampling rate is $250\,{\rm MS/s}$. We obtain one data point of the IQ quadratures in each $16\,{\rm ns}$ period, which is then digitally filtered by a $500\,{\rm kHz}$ low-pass finite-impulse-response filter before averaging. The repetition rate of the measurement is $1.25\,{\rm kHz}$. In each repetition, we obtain a $32\,{\rm \mu s}$-long trace of the IQ quadratures, which is subsequently averaged by $2\times 10^{4}$ times.

\subsection{Averaged response.}
The instantaneous reflection coefficient of the thermometer is described as $S_{11} = 1 - \textrm{e}^{\textrm{i}\phi}\gamma_{\rm c}/\left[(\gamma/2) + \textrm{i}\Delta\right]$\,\cite{Chen2022b}, where $\gamma_{\rm c}$ and $\gamma$ are the external and the total energy decay rates, and $\Delta=2\pi(f_{\rm r}-f_{\rm p})$ is the detuning between the resonance and the probe frequencies. The parameter $\phi$ describes the asymmetry of the resonance. We consider a Gaussian distribution of the resonance frequency, $f_{\rm r} \sim \mathcal{N}\left(\mu, \sigma^2\right)$, which is caused by the statistical properties of the input photons (see Supplementary Note 2). The averaged thermometer response is thus given by
\begin{align}
    S_{11}^{\rm ave} = 1 - \frac{\textrm{e}^{\textrm{i}\phi}\gamma_{\rm c}}{2\sqrt{2\pi} \sigma} 
    {\rm erfcx}\left(\frac{(\gamma/2) + \textrm{i}\Delta'}{2\sqrt{2}\pi \sigma}\right).
\end{align}
Here, $\Delta'=2\pi(\mu-f_{\rm p})$ and ${\rm erfcx}(\cdot)$ is the scaled complementary error function (see Supplementary Note 3). The magnitude of $S_{11}^{\rm ave}$ is equivalent to the Voigt profile in laser spectroscopy\,\cite{Demtroeder1996}. Considering the contribution of the external circuitry, the real measured reflection coefficient is denoted as $S_{11}^{\rm raw}$ (see Supplementary Note 4).

\section*{Data availability}
The data that support the findings of this study are provided in the paper. Source data are provided with this paper.  

\section*{Code availability}
The codes for analyzing the data of this study are provided with this paper. 

\section*{References}
\bibliographystyle{naturemag}
\bibliography{BoloG2_REF.bib} 

\section*{Acknowledgments}
\noindent We thank Matti Partanen and Kirill G. Fedorov for the preliminary characterization of the nanobolometer. 
This work is supported by the Academy of Finland Centre of Excellence program (No.\,$336810$), European Research Council under Advanced Grant ConceptQ (No.\,$101053801$),  Business Finland Foundation through Quantum Technologies Industrial (QuTI) project (No.\,41419/31/2020), Technology Industries of Finland Centennial Foundation, Jane and Aatos Erkko Foundation through Future Makers program, Finnish Foundation for Technology Promotion (No.\,$8640$), and Horizon Europe programme HORIZON-CL4-2022-QUANTUM-01-SGA via the project 101113946 OpenSuperQPlus100.

\section*{Author Contributions Statement}
A.K. and Q.C. performed the experiment and analyzed data. 
J.G. fabricated the nanobolometer.
Q.C. developed the thermal radiation source.
Q.C. and A.K. programmed FPGA and developed measurement software.
A.G. contributed to measurement software.
V.V. contribute to experimental setup.
P.S., J.M., and Q.C. contributed to nanobolometer design and fabrication. 
Q.C., A.K., and M.M. wrote the manuscript with input from all authors. 
M.M. contributed to data analysis and supervised the project. 

\section*{Competing Interests Statement}
M.M. is a co-founder and shareholder of IQM\,Finland\,Oy. M.M. is an inventor of patents FI122887B, US9255839B2, JP5973445B2, and EP2619813B1 titled "Detector of single microwave photons propagating in a guide". Other authors declare no competing interests.

\newpage
\begin{figure}
  \centering
  \includegraphics[width=9cm]{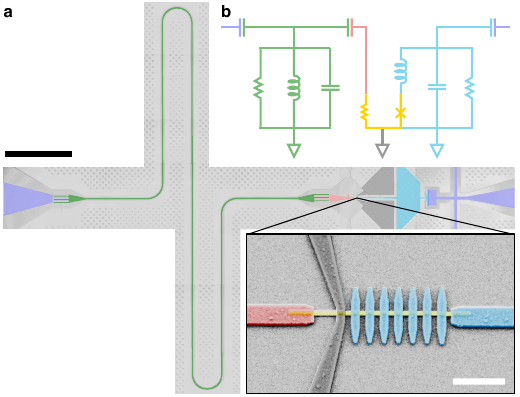}
  \linespread{1.2}
  \caption{{\bf Photograph of sample and equivalent circuit.} 
{\bf a} False colored optical image of a reference sample. The circuit elements are color coded as launch pad (purple), transmission-line filter (green), absorber (red), ground line (gray), and thermometer (blue). The normal-metal nanowire across the absorber and thermometer is colored in yellow. The scale bar denotes $500\,{\rm \mu m}$.
The inset shows a scanning electron microscopy (SEM) micrograph of the long- and short-SNS junctions of the nanobolometer, where the scale bar denotes $2\,{\rm \mu m}$.
{\bf b} Equivalent lumped-element circuit of the system with the same color code as in (a).
}
\end{figure}

\begin{figure}
  \centering
  \includegraphics[width=18cm]{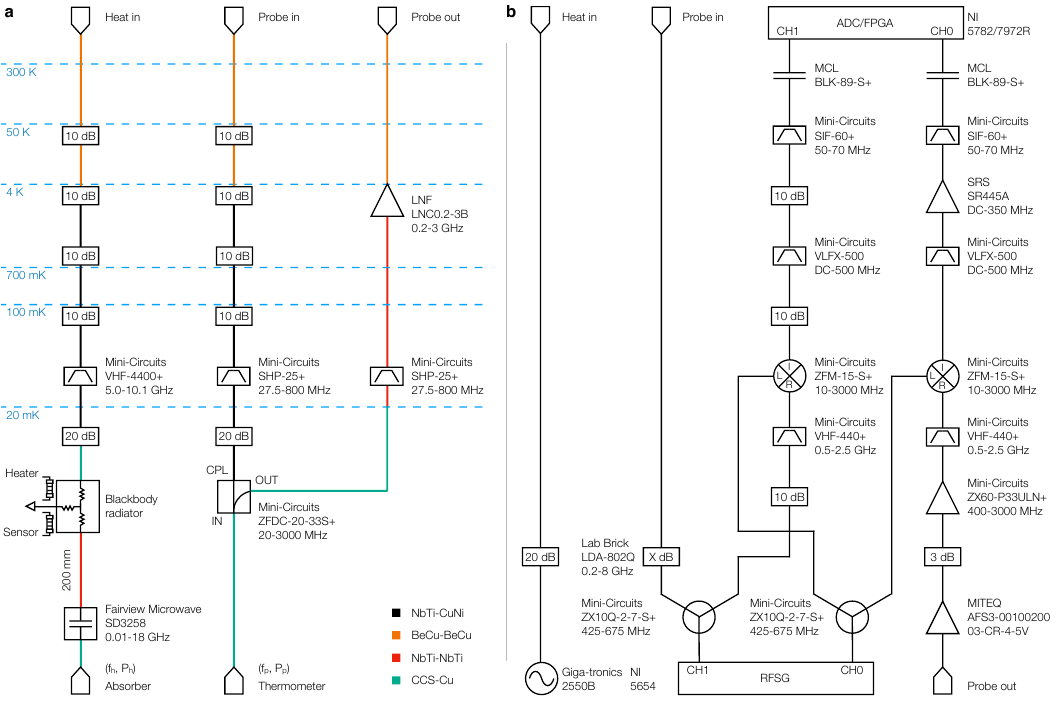}
  \linespread{1.2}
  \caption{{\bf The detailed experimental setup.} 
{\bf a} The sample is mounted at the MXC plate of a Bluefors LD250 dilution refrigerator. Signals are routed between the cryogenic and room temperatures via coaxial microwave cables with different materials along with different microwave components for maximizing the signal-to-noise ratio. The blackbody radiator is suspended below the MXC plate and is weakly thermalized to the latter for local temperature control.
{\bf b} The room-temperature setup for photon moments measurement. The probe-out signal is amplified and down-converted for data acquisition, while the probe-in signal is also recorded simultaneously for phase reference.
}
\end{figure}

\end{document}